# REDEFINING THE BOUNDARIES
# OF INTERPLANETARY CORONAL MASS EJECTIONS
# FROM OBSERVATIONS AT THE ECLIPTIC PLANE

C. Cid[1], J. Palacios[1], E. Saiz[1] and A. Guerrero[1]

[1] Space Research Group - Space Weather, Departamento de Física y Matemáticas, Universidad de Alcalá, Alcalá de Henares, Spain.

# ABSTRACT

On 2015 January 6-7, an interplanetary coronal mass ejection (ICME) was observed at L1. This event, which can be associated with a weak and slow coronal mass ejection, allows us to discuss on the differences between the boundaries of the magnetic cloud and the compositional boundaries. A fast stream from a solar coronal hole surrounding this ICME offers a unique opportunity to check the boundaries' process definition and to explain differences between them. Using Wind and ACE data, we perform a complementary analysis involving compositional, magnetic, and kinematic observations providing relevant information regarding the evolution of the ICME as travelling away from the Sun. We propose erosion, at least at the front boundary of the ICME, as the main reason for the difference between the boundaries, and compositional signatures as the most precise diagnostic tool for the boundaries of ICMEs.

## 1. INTRODUCTION

Coronal Mass Ejections (CMEs) are intense phenomena from the solar upper atmosphere that have impact through the entire heliosphere, often triggering geomagnetic storms when they interact with the terrestrial magnetosphere, making them highly significant in space weather studies. CME plasma is initially confined in closed magnetic loops, which are opened up by reconnection (e.g., Fisk 2003; Antiochos et al. 2011).

The interplanetary manifestations of CMEs, frequently labelled as interplanetary coronal mass ejections (ICMEs), are characterized by a collection of signatures which includes, among others, magnetic field, plasma dynamics and composition signatures (Zurbuchen & Richardson 2006, and references therein). A subset of ICMEs characterized by enhanced magnetic fields with smooth rotation, and low temperature (Burlaga et al. 1982; Burlaga 1991) is termed magnetic clouds (MCs) and a flux-rope structure is commonly associated with these ejections.

Rust (1994) and Bothmer & Schwenn (1994) showed that the chirality (sign of the magnetic helicity) of MCs associated with filament eruptions from the north (south) hemisphere is roughly negative (positive). Rust (1994) and Low (1996) used magnetic helicity conservation in conceptual models linking helicity spawned by the dynamo with transfer of helicity to the corona, where it would accumulate and it would be eventually released into interplanetary space through CMEs. Thus, chirality in a flux rope should be conserved when travelling from the Sun to the Earth.

The first studies of ICME identification and evolution through the heliosphere (e.g. Burlaga et al. 1981) deal with MCs as they are relatively easy to identify over the background solar wind. The fast-mode interplanetary shock that often precedes them has also been a common way to identify ICMEs in the inner heliosphere and to trace their effects outward (e.g. Wang & Richardson 2001). Far from the Earth and out of the ecliptic plane, some compositional signatures, such as the helium abundance enhancement, have shown to be relevant in the identification of ICMEs (Paularena et al. 2001; Von Steiger & Richardson 2006; Wimmer-Schweingrubber et al. 2006).

The different plasma dynamics in and out of the ecliptic plane was an encouraging aspect in the study of solar wind plasma composition anomalies to identify ICMEs, since the ionic charge composition of the solar wind is expected to become frozen-in at several solar radii, reflecting the electron temperature in the corona (Hundhausen 1968). The oxygen charge state ratio $O^{7+}/O^{6+}$ has long been used as a good representative magnitude of the solar wind source region because of its relatively fast freeze-in process in the low corona (Bürgi & Geiss 1986). A low $O^{7+}/O^{6+}$ ratio is considered to be a very good signature of a stream from a coronal hole (Zurbuchen et al. 2002), while periods of unusually hot charge states, with high $O^{7+}/O^{6+}$ ratio, are typically associated with CMEs (Lepri et al. 2001; Henke et al. 2001). In contrast, the average Fe charge state ($<Q_{Fe}>$) has been shown to be a sensitive tracer of electron temperatures at large solar heights (up to 4 $R_S$), so that it can be used as a measure of the evolutionary properties in the far corona (e.g., Lepri et al. 2001; Lepri & Zurbuchen 2004; Gruesbeck et al. 2011). For this reason, high Fe charge states ($Q_{Fe}$) are an excellent and sufficient signature for identifying identify ICMEs (Lepri et al. 2001; Lepri & Zurbuchen 2004).

Generally, ICMEs also contain an enhanced ratio of $He^{2+}/H^+$. Although it is far from clear exactly how and where such a helium accumulation is produced and how it can persist for any time interval until it is ejected with a CME, there is a large body of literature regarding abundance variations of He in the corona. For example, Lepri et al. (2013) and references therein suggest that lower helium abundances are a result of less efficient Coulomb drag.

The boundary between an ICME and ambient solar wind might be expected to be a discontinuity that encompasses a region with ICME-like signatures. In practice, the various ICME signatures do not indicate exactly the same boundaries (Wimmer-Schweingruber et al. 2006; Zurbuchen & Richardson 2006; Richardson & Cane 2004)). Moreover, a region such as that considered in this paper may have distinct boundaries in plasma, magnetic field and/or other signatures, and its boundaries may be ambiguous. This is a relevant issue for theoretical flux-rope models, which fit magnetic data inside an ICME in order to obtain the MC axis; therefore the results are highly dependent on the selected boundaries. There are also ICMEs that lack some of the characteristic signatures; hence, to date, a clear characterization that defines the presence of an ICME and its boundaries remains elusive.

The aim of this paper is to perform an analysis of solar and solar wind data for the ICME event on 2015 January 6-8. We place emphasis on looking for ICME boundaries, taking into account different parameters and comparing ICME composition and kinetic properties. We introduce the interplanetary data used in this study in Section 2. We first analyze the magnetic and plasma signatures, and then the compositional signatures, focusing on the possible boundaries of the ICME existing in the interval. Then, in Section 3, we examine the solar activity by searching for the precursors of the interplanetary transients. In Section 4 we discuss the findings of these analyses and their implications in the overall picture of the definition of ICME boundaries. Conclusions are presented in Section 5.

## 2. INTERPLANETARY OBSERVATIONS

In the time interval to be discussed, 2015 January 5-9, in situ physical magnitudes are obtained from different instruments on board Wind and ACE. At that time, both satellites were located upstream of the Earth at the L1 point. We obtained the magnetic field data from the Wind/MFI (Lepping et al. 1995) and ACE/MAG (Smith et al. 1998), the plasma data from the Wind/SWE (Ogilvie et al. 1995) and ACE/SWEPAM (McComas et al. 1998), and ionic composition from ACE/SWICS (Gloeckler et al. 1998). The time resolution ranged from 1 minute to 2 hours for different data sets.

### 2.1. Magnetic and Plasma Signatures

The basic magnetic field and plasma observations from the Wind spacecraft are shown in Figure 1. Similar observations were recorded (not shown) by ACE, but had some data gaps at the times of large density values.

An MC (Burlaga et al. 1981; Klein & Burlaga 1982) appears on January 7, which can be identified (shadowed area in Figure 1) by the relative high magnetic field strength $B$, the smooth rotation of the elevation angle $\theta$ and the low proton temperature ($T$). We have established the MC boundaries at 07:50 UT and 19:45 UT on January 7 considering the smoothness of the magnetic field vector. The proton temperature is low in this interval; however, taking into account the low-temperature criterion the front boundary of the MC could be shifted ahead of the current position. The smooth rotation of interplanetary magnetic field from south to north along with positive $x$ and $y$ GSM-components (not shown), indicates that the helicity of the MC is left-handed.

The MC is travelling at a similar velocity to the surrounding solar wind, which is around 500 km s$^{-1}$. A discontinuity in $B$ at 5:39 UT on January 7 (dashed line labelled as SL in Figure 1) precedes the MC and similar changes can be also appreciated in solar wind velocity and density, although no discontinuity appears in temperature. Therefore, SL will not be labelled as an interplanetary shock, but only as a shock-like discontinuity.

Other discontinuities appear in magnetic field strength, dynamic pressure or temperature, but only two of them appear simultaneously at the three plasma parameters: FS corresponds to a forward shock and RS to a reverse shock, indicating the expansion of the material in between (note the enhancement of solar wind density).

Ahead FS and behind RS, the proton density is low (less than 5 cm$^{-3}$), the solar wind velocity is fast (close to 500 km s$^{-1}$), and the temperature is high (~ 2×10$^5$ K). These signatures correspond to a fast stream from a solar coronal hole.

## 2.2. Compositional signatures

Figure 2 shows a summary of compositional parameters for the same period as in Figure 1. The shadowed area and vertical dashed lines appear at the same times as those in Figure 1 for comparison purposes. Magnetic field strength and solar wind density from Wind/SWE are shown again in panels (a) and (b).

The values of He$^{2+}$/H$^+$ from Wind/SWE appear in panel (c) of Figure 2. The horizontal dashed line in this panel indicates He$^{2+}$/H$^+$ = 0.06. Following Richardson & Cane (2004), we use the criterion He$^{2+}$/H$^+$ ≥ 0.06 to identify enhanced helium abundances typically associated with interplanetary ICMEs. In the analyzed interval, we find at the rear boundary of the MC (shadowed area) a two-hour interval with enhanced helium abundance well exceeding the threshold.

Panels (d) and (e) of Figure 2 show the average Fe charge state and the O$^{7+}$/O$^{6+}$ ratio respectively at a 2-hour cadence measured by ACE/SWICS. Superimposed on the observed values, as gray solid lines, are the expected values inferred from the simultaneously observed solar wind, according to Table 1 of Richardson & Cane (2004). The $O^{7+}/O^{6+}$ ratio is higher than the expected values for ambient solar wind between 11:30 UT (January 6) and 05:30 UT (January 8). This time interval, which is more than three times longer than that between the MC boundaries, coincides with that of the low solar wind temperature, indicating a larger interval for the ICME material. The time interval corresponding to higher values than the expected for the average Fe charge state is shorter than that for the $O^{7+}/O^{6+}$ ratio. Nevertheless, the interval for the $O^{7+}/O^{6+}$ ratio coincides with that of the average Fe charge state above 10. According to Lepri & Zurbuchen (2004), the probability of finding charge states higher than 10 in the normal solar wind rapidly drops off.

Proton specific entropy, as well as solar wind composition, can be used to identify plasmas of different origin and have been recognized as marker of boundaries between different types of solar wind (Burlaga 1990, Wimmer-Schweingruber et al. 1997, Pagel et al. 2004). Following prior work by Burton et al. (1999) and Pagel et al. (2004), we have computed $ln(T_p/N_p^{\gamma-1})$, as a proportional quantity to proton specific entropy, where $\gamma$=1.5, and $T_p$ and $N_p$ represent the proton temperature and number density, respectively (see (f) panel in Figure 2). The proton specific entropy and the $O^{7+}/O^{6+}$ ratio show good correlation throughout the shown interval, according to the results of Pagel et al. (2004) for some periods of low solar activity. Indeed, between FS and RS the proton specific entropy decreases, these two lines appearing as the boundaries between the fast solar wind from the coronal hole and the ICME material. According to this scenario, forward-reverse (F-R) shocks correspond to the boundaries of the ICME material.

## 3. SOLAR OBSERVATIONS

To study the solar source of this event we have used data from SDO (Pesnell et al. 2012), more particularly, from AIA (Lemen et al. 2012) 304 and 193 Å, 12-min cadence, from January 2 00:00 UT to January 5 00:00 UT (i.e., about five and two days before the arrival of the MC at L1). For completeness, Hα processed images from the GONG[1] observatories (http://halpha.nso.edu/) (Harvey et al. 2011) have been used for comparison with AIA 304 Å for the same period and variable cadence due to gaps (GONG theoretical intended cadence: 1 min). We also estimate the chirality of the filaments and structures by checking SDO/HMI line-of-sight magnetograms (Scherrer et al. 2012).

---

[1]Now part of the NSO Integrated Synoptic Program (NISP) http://nisp.nso.edu/aboutnisp.

In addition to these datasets, we have also used data from the EUV imaging telescope SWAP (Seaton et al. 2013, Halain et. al 2013) 174 Å, with cadence of 1-2 minutes, and spatial sampling of 3.16" px$^{-1}$. The already reduced dataset used here is level 1 on reduction pipeline version v1.5. SWAP is onboard PROBA2 (Santandrea et al. 2013). For CME candidate searching and corona visualization, we have used LASCO C2 images (Brueckner et al. 1995) onboard SOHO (Domingo et al. 1995).

Figure 3 displays a context Hα image, which will be helpful in following the description of the abundant solar activity occurring during January 2-4. In the figure, the main active regions (ARs) are marked in pale yellow, the main coronal hole (with negative polarity) in red and filaments exhibiting sinistral chirality in black and dextral in blue. First we describe the CME candidates and then the role of the active filaments.

Coronal activity is as follows in this period. On January 2, at 14:36 UT, there was a limb CME observed by LASCO in the west limb. The estimated plane-of-the-sky (POS) speed is about 300 km s$^{-1}$. The mean POS speed calculated by CACTUS (Robbrecht & Berghmans 2004; Robbrecht, et al. 2009) is 312 km s$^{-1}$. This CME seems to be associated with a limb prominence that was totally peeled off at 14:24 UT in the AIA 304 Å data. However, the CME origin may be more disk centered, as some dimmings are observed about one hour before the eruption at AR 12256 and north of the equator. Moreover, the role of the AR 12256 emergence in the magnetic instability of this region cannot be discounted, since this area exhibits several brightenings during the whole period, as a C1.2 flare at 19:48 UT.

The other CME candidate is the following. At 03:12 UT on January 3, a CME ejected from the south limb makes its first appearance in the LASCO/C2. The ejection is faint and slow, but visible. The filament and cavity structures are not discernible. The CME visual angular width is more than 100º. The estimated plane-of-the-sky (POS) speed is 120 km s$^{-1}$. CACTUS[2] mean POS speed is 134 km s$^{-1}$. LASCO catalog[3] also

---

detected it, with a POS speed of 160 km s$^{-1}$. At CORIMP catalog[4] (Byrne et al. 2012; Morgan et al. 2012), the CME is observed as clearly driven by the fast wind of the southern coronal hole. This CME is displayed in Figure 4, showing a base difference in logarithmic scale.

The coronal ejection is associated with a dimming enlargement observed in AIA 193 and 304 Å and Proba-2/SWAP close to the southern coronal hole (see Figure 5). During January 2, there is an area of reduced density compared to the surroundings in the north-east boundary of the southern coronal hole. Close to midnight of January 2, the dimming started increasing in area. Then, on January 3 at 07:24 UT there was an extended and massive but weak ribbon flare in the CH boundaries, with a large area increase during the slow ejection (see Figure 5, mainly at coordinates [-200", 400"] but also at the whole coronal hole boundary). The west part of the coronal hole boundary flared a little later, at 08 UT. Coincidentally, also the north of AR 12254 is flaring in AIA 304 Å. After this flare-like event and ejection, the CH boundaries slowly reduced. In Figure 6 we show the measured area of the coronal hole from January 2, 00:00 UT to January 5, 00:00 UT using a Lambertian (i.e. equal-area) map projection. The area increase was from $1.56 \times 10^{11}$ km$^2$ to $1.93 \times 10^{11}$ km$^2$, that is, a 23.5%.

Now we describe the state of the chromosphere and low corona through data from SDO/AIA, Hα from NISP and Proba-2/SWAP, focusing on filament activity. There are a number of occurrences observed in Hα. Different destabilizations, oscillations and disappearances in Hα (these may be due to any kind of thermal disturbance than makes the emission in Hα lessen or lose contrast due to a background increase; i.e, a flare) may be the signature of a real eruption of a filament, but in the cases described below, apparently there is no eruption of a filament unless stated as such. These filaments are defined as critical as they seem to have destabilized at some point in this two-day period.

The filament F1 is one of the most active, as it exhibits different important destabilizations at 03, 05 UT (such as expanding and losing its structure) and 16 UT on January 2.

---

[4]http://alshamess.ifa.hawaii.edu/CORIMP/realtime/soho/lasco/detections/2015/01/02/cme_ims_orig_20150102_050606/movie_C3.html

The filament F2 starts becoming visible on January 2, 07 UT. It is very close to F1 and actually in interaction some time after. On January 3, 06:30 UT, F1 and F2 filaments start to merge, showing a unique structure around 15 UT in 304A. This merging may have ejected a piece of filament with sinistral chirality as a by-product.

The filament F3 in AR 12254 is small, but displays activity due to several brightenings. For instance, a brightening at 03:12 UT affects the filament but after this, the filament seems to be enlarged. This filament is also a candidate for eruption at the time of the ribbon flare.

Another critical filament, as probably erupting, is F4, embedded in AR 12253. This filament weakened due to a flare on January 3 at 06:29 UT, close to the time of the ribbon flare. From 05:52 to 07:05 UT there were five C-class flares, due to the extension (and cancelation of some positive adjacent polarities) of the negative intrusion into the large positive sunspot. After this period, the filament reappeared, deformed and grew, probably due to the underlying flux emergence and rearrangement. At 09:40 UT, a M1.1 flare occurred in AR 12253, one of the largest in this period. Twice thereafter, F4 as observed in Hα disappeared, but it subsequently appeared restructured.

The most active region is AR 12253, with several flares during the analyzed period. There are several short filaments in this region, and it is difficult to assess the possible eruption of these filaments and their relationship with the flares. Due to the activity on AR 12253, the filaments are barely visible in Hα from 00:00 UT to 09:30 UT on January 4. The largest registered flare in the interval analyzed was an M1.3 on AR 12253 on January 4 at 15:35 UT. A piece of filament erupted and then disappeared at 18 UT, but it subsequently rearranged.

AR 12253 exhibits an intrusion of a negative polarity region into the positive preceding spot, and this fact implies a larger amount of flares and it may translate into filament eruptions. This protrusion is similar to the one described in Toriumi et al. (2013), which produced strong flaring activity and in the end it involved filament eruption.

The filament F5 at AR 12252 is one of the most motile structures, since it suffers conspicuous oscillations as observed in Hα during January 2, as a vigorous longitudinal oscillation from 16 to 17 UT. On January 3, at 16:25 UT, F5 destabilized and disappeared forever, as it apparently erupted. The role of this filament can be ruled out, since its chirality is dextral.

In addition to all this activity, on January 4 14:30 UT, a filament of AR 12248 (bottom-right corner of the yellow box in Figure 3) was ejected, with some previous brightenings at 03:30 and 09 UT. Its chirality is sinistral.

The rest of the filaments shown in the image are apparently stable in this two-day period.

## 4. DISCUSSION

In Section 2 we have described several signatures in the solar wind observations, which indicate that an ICME with a flux-rope topology passed the Wind and ACE spacecraft on 2015 January 6-7. The most probable solar source of this disturbance is a CME that reached the LASCO C2 field of view at 03:12 UT on January 3. It was ejected from the south limb with several non-clearly associated on-disk signatures, and without major associated solar flares. The speed of the ICME is consistent with the arrival time late on January 6. However, a projection effect in CME POS velocity (slightly above 100 km s$^{-1}$) is not enough to reach the speed of the ICME, which is probably increased due to the acceleration of the CME by the surrounding fast solar wind.

There was another CME in the interval analyzed at 14:36 UT on January 2, triggered by a limb prominence. The low probability for a limb prominence to be observed as a flux rope at L1 point leads us to discard it as the solar source.

We define the MC front and rear boundaries at Wind at 07:50 UT and 19:45 UT (January 7). However, these boundaries do not coincide with any discontinuity in composition or other solar wind plasma parameters such as temperature, but with two large enhancements in proton number density. A sharpened discontinuity between the solar wind and the ICME material appears in the $O^{7+}/O^{6+}$ ratio. According to this parameter, we should set the ICME boundaries at 11:30 UT (January 6) and 05:30 UT (January 8). Between these boundaries, also high Fe charge states are also observed, supporting the limits of ICME material. The fact that no major flares occur at the time of the solar ejection may be the reason that the Fe charge states do not reach larger ionization values. Another plausible explanation for these observed average Fe charge states between the boundaries could be the presence of some bi-modal material with a nearly ubiquitous presence of the ICME plasma with a high Fe charge state peak (~ $Fe^{+16}$) and a more typical solar wind (~ $Fe^{+10}$) (Gruesbeck et al. 2011). An F-R shock pair, marked by vertical dashed lines in Figures 1 and 2, bounds the ICME region as defined by compositional boundaries. Gosling et al. (1994) claimed that they discovered, out of the ecliptic plane, a new class of ICMEs fully embedded in polar fast streams producing an F-R shock pair in the solar wind. This F-R shock pair was associated with the expansion of the ICME because of its high internal dynamic pressure when compared to the surrounding solar wind. Similar conditions to those presented by Gosling et al. (1994) appear in the analyzed interval. Indeed, as in Ulysses observations, the ICME analyzed in this paper shows a higher internal dynamic pressure than the bounding fast streams where it is embedded (see panel (f) in Figure 1).

As described above, the ICME region is compositionally distinct from the rest of the solar wind, indicating that it likely originates at a different height in the solar corona and it is accelerated by a different mechanism. On the other hand, the F-R shock pair bounding the ICME indicates that this material undergoes a common expansion history. Therefore, if magnetic topology and kinematic properties are different perspectives of the same plasma, why are magnetic boundaries so distinct from compositional ones?

As the ICME propagates away from the Sun, it interacts with the ambient solar wind. This interaction can include changes in the magnetic connectivity when, under a favorable orientation, magnetic reconnection between the ICME and the ambient solar wind takes place. In that case, reconnection peels off magnetic flux from the leading edge of the ICME in a process labelled as erosion (Dasso et al. 2006, 2007, Ruffenach et al. 2012, 2015), which may also modify the topology of the field lines at the trailing part of the ICME. Figure 7 illustrates the flux rope as observed by Wind during its propagation to the Earth. As deduced from the observations, at the time of the encounter with the spacecraft, the IMF $B_z$ component rotates from south to north and the axis of the MC is pointing not far from the dawn direction (IMF strength is mainly due to a positive $y$ GSM-component). The yellow lines in Figure 7 represent the fast solar wind flow from the southern solar coronal hole with negative magnetic polarity. This fast wind flow, in which the MC is embedded, provides favorable conditions for erosion. As a result, a large amount of IMF pointing southward in the front boundary of the cloud disappears. It is important to keep in mind this issue because this erosion process may lead to in an incorrect determination of the actual axis orientation of the flux rope when fitting theoretical models to observations.

From 16:19 UT (January 6) until the SL discontinuity at 05:39 UT (January 7), which is only two hours ahead the front boundary of the MC, the interplanetary magnetic field is northern or close to zero (see panel (b) in Figure 1). These observations can be considered as the remnant of the reconnection between a fluctuating field from the fast stream preceding the ICME and the front part of the ICME, which can be assumed as pointing southward according to the south-to-north rotation of the MC. This evolution would also explain the decrease of the specific entropy when approaching from the fast stream to the MC, which is the interval with lower proton specific entropy (see panel (f) in Figure 2).

A step-like discontinuity appears in the specific entropy at the rear boundary of the MC that draws our attention. This interval also presents an enhancement in proton number density. Furthermore, it coincides with the two-hour interval with enhanced helium to proton number density ratio well exceeding the threshold. Indeed, the helium density exceeds by 10 % the proton number density, consistent with the photospheric abundance of $He^{2+}$. Considering the similarity of these features with those described by Burlaga et al. (1998) and Kozyra et al. (2013), we suggest that this high-density region at the rear of the magnetic cloud might be filament material.

Solar observations provide several possibilities for the source of the filament material, but unfortunately, they are not conclusive. As the typical three-part structure cannot be distinguished in these LASCO images for the CME on January 3, it is hard to identify unequivocally the filament material with its source on the solar disk. Taking into account the left-handed helicity of the MC, the close correspondence between features in the solar wind and solar observations indicates a sinistral chirality for the solar filament. This is the case for some of the filament candidates, being located in AR 12251 and AR 12253, and even 12254. The filamentary structure located east of the coronal hole (close to the quiescent filament) cannot be ruled out either. The quiescent filament and the filamentary structure are located on X = [-800,-200], Y = -500 arcsec in Figure 5. The ear-shaped filament F5 in AR 12252, albeit very active, is dextral, so it is discarded as a candidate.

## 5. CONCLUSIONS

The present work provides important new information regarding the definition of the boundaries of ICMEs and MCs. The main findings of the present study can be summarized as follows:

1. Compositional signatures are precise diagnostic tools for the boundaries of ICMEs. In the analyzed event, the step in the $O^{7+}/O^{6+}$ ratio indicates how well the stream interface has been conserved between plasmas of different composition during the transit of almost four days from the event at 2 $R_S$ to 1 AU. The ICME region also presents high Fe charge states, but it is no higher than the expected values inferred from the simultaneously observed solar wind during the whole interval. Case studies, such as the one analyzed in this paper, may be useful to revise the expected values obtained through statistical studies, or at least their uncertainty.

2. Kinematic analysis of the event reveals an F-R shock pair at the boundaries of the ICME. This kind of shock pair, commonly observed at high latitudes, is not typical at 1 AU at the ecliptic plane, due to the high speed of the ICMEs relative to the 'usual' solar wind. In this case, the ICME presents a similar velocity to (or even lower than) that of the surrounding solar wind, providing a scenario which can be compared to that at high latitudes, where ICMEs are slower than the fast solar wind from coronal holes.

3. Our boundary identification from magnetic signatures does not agree with those from other signatures. This discrepancy is explained as due to solar wind interaction, which modifies magnetic topology of ICMEs. Complementary analyses involving compositional, magnetic, and kinematic observations can provide relevant information regarding the evolution of the ICME while travelling away from the Sun. This is the case in the present paper, where erosion at least at the front boundary of the ICME produces a disappearance of a large part of the southern component in the MC. This interaction may result in inaccurate estimations of the flux-rope axis when not considered in theoretical models, and also in a decrease of the expected geoeffectiveness predicted from solar observations.

**Acknowledgements**. This work was supported by MINECO project AYA2013-47735-P. We acknowledge the use of publicly available data products from Wind/MFI

Wind/SWE, ACE/MAG, ACE/SWEPAM, ACE/SWICS, SDO/AIA, SDO/HMI, SOHO/ LASCO and Hα processed images from the GONG/NISP observatories. SWAP is a project of the Centre Spatial de Liege and the Royal Observatory of Belgium funded by the Belgian Federal Science Policy Office (BELSPO). LASCO CME catalog is generated and maintained at the CDAW Data Center by NASA and The Catholic University of America in cooperation with the Naval Research Laboratory. SOHO is a project of international cooperation between ESA and NASA.

**Figure Captions**

**Figure 1.** Solar wind magnetic field and plasma parameters from Wind during the interval 2015 January 5.5-9 at 1-minute resolution. The shadowed interval denotes a magnetic cloud. Vertical dashed lines correspond to a forward shock (FS), a shock-like discontinuity (SL) and a reverse shock (RS).

**Figure 2.** Magnetic field strength (a), proton density (b) and composition parameters (c) - (e) from Wind during the interval 2015 January 5.5-9. Panel (c) shows values of $He^{2+}/H^+$ density ratio at 1-minute resolution from Wind/SWE. Panels (d) and (e) display the average Fe charge state and the $O^{7+}/O^{6+}$ ratio, respectively, at a 2-hour cadence from ACE/SWICS. Panel (f) shows $ln(T_p/N_p^{\gamma-1})$, as a proportional quantity to proton specific entropy for the same interval. The shadowed interval and vertical dashed lines are the same as in Figure 1. The horizontal dashed line in the panel (c) indicates $He^{2+}/H^+ = 0.06$. Panels (d) and (e) show, in gray solid line, the expected values inferred from the simultaneously observed solar wind appear (see text for details).

**Figure 3.** Context image in Hα marking the most prominent features: ARs in yellow, an emerging AR in turquoise, sinistral filaments in black and dextral filaments in blue. A red oval indicates the position of a southern coronal hole, not visible in Hα.

**Figure 4.** LASCO base difference image in logarithmic scale showing the CME in the south-east sector at 09:12 UT on January 3. The CME is depicted in a lighter shade of red.

**Figure 5.** The sequence of images showing the enlarged dimming due to the CME. In the three rows, SDO/AIA 304 Å are shown on top, SDO/AIA 193 Å in the middle and Proba-2/SWAP 174 Å at the bottom. In the left panels, an area of reduced density in the North-east is visible at 00:00 UT. At 07:00 UT, a flare surrounded the whole CH boundary. After that, at 11:00 UT, the dimming is evident (right panels).

**Figure 6.** Plot of the CH area vs. time. The increasing area of the CH corresponds to the dimming. The time interval starts on January 2, 00:00 UT and the maximum area of the dimming is at 10:48 UT of the next day.

**Figure 7.** A simplified sketch illustrating the propagation of the ICME (flux rope in red) bounded by a fast stream from the coronal hole (yellow lines) close to the arrival to the L1 point. The formation of an "X-point" appears in the front boundary of the flux rope where magnetic reconnection between the ICME and the ambient solar wind takes place.

FIGURE 1

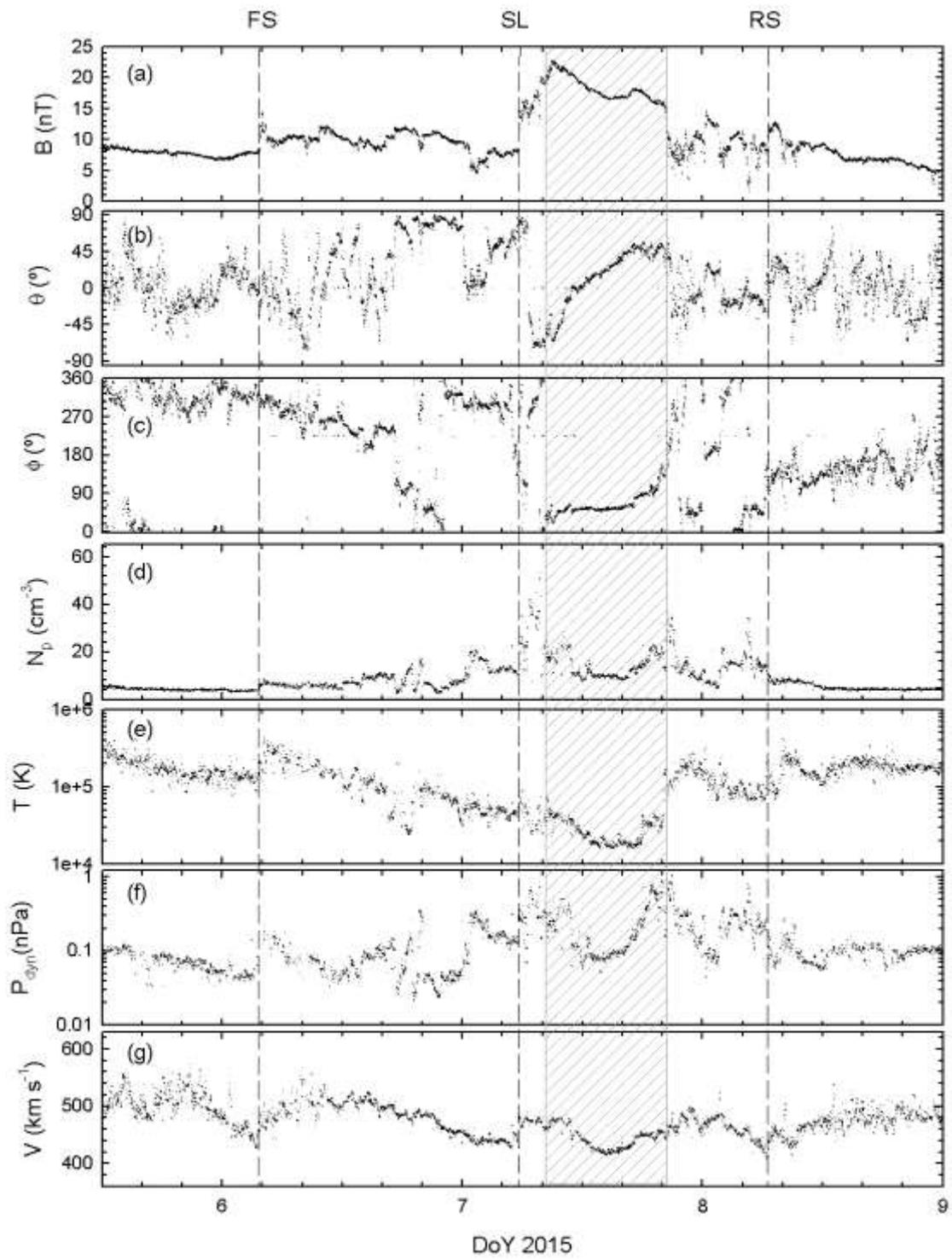

FIGURE 2

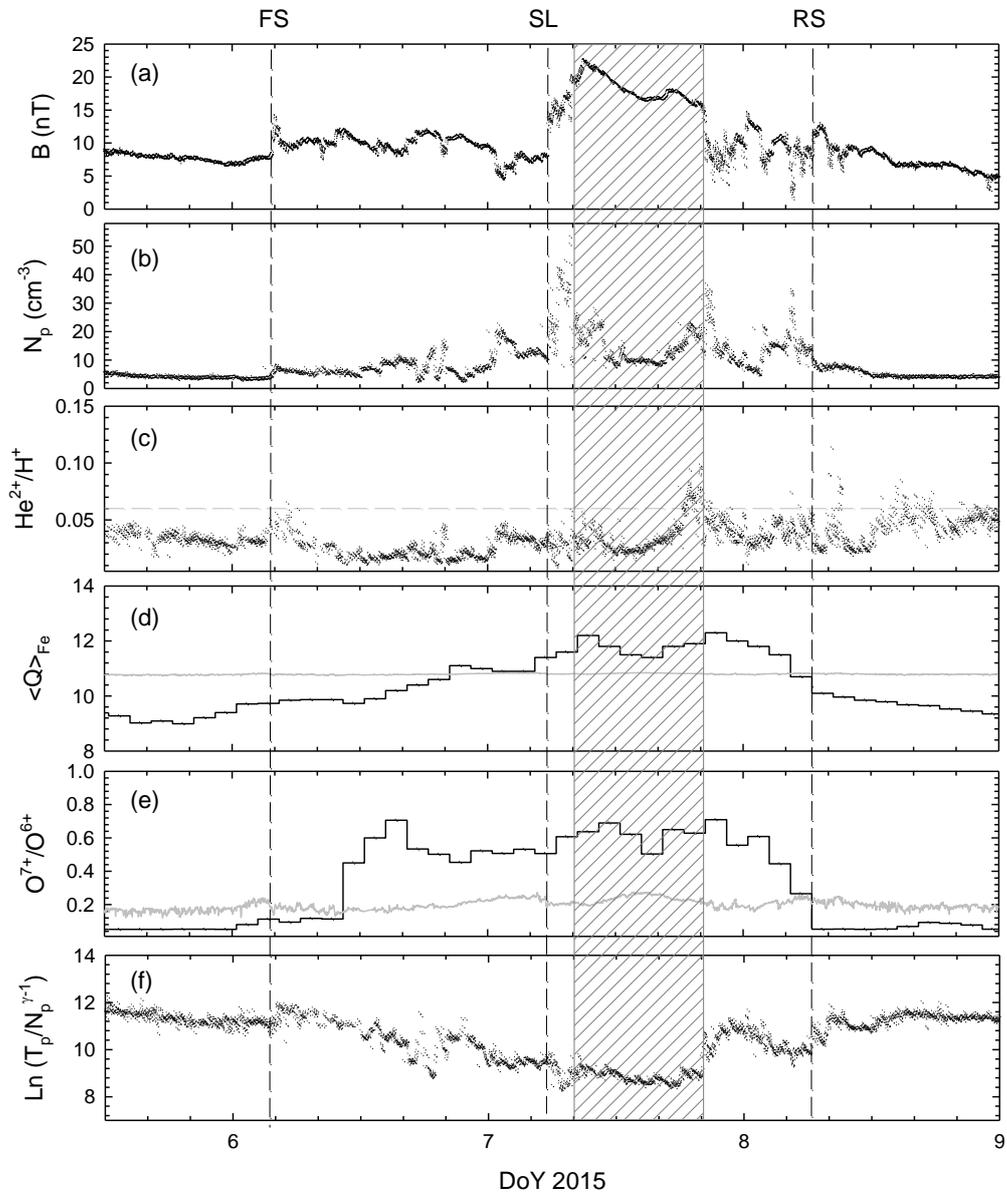

FIGURE 3

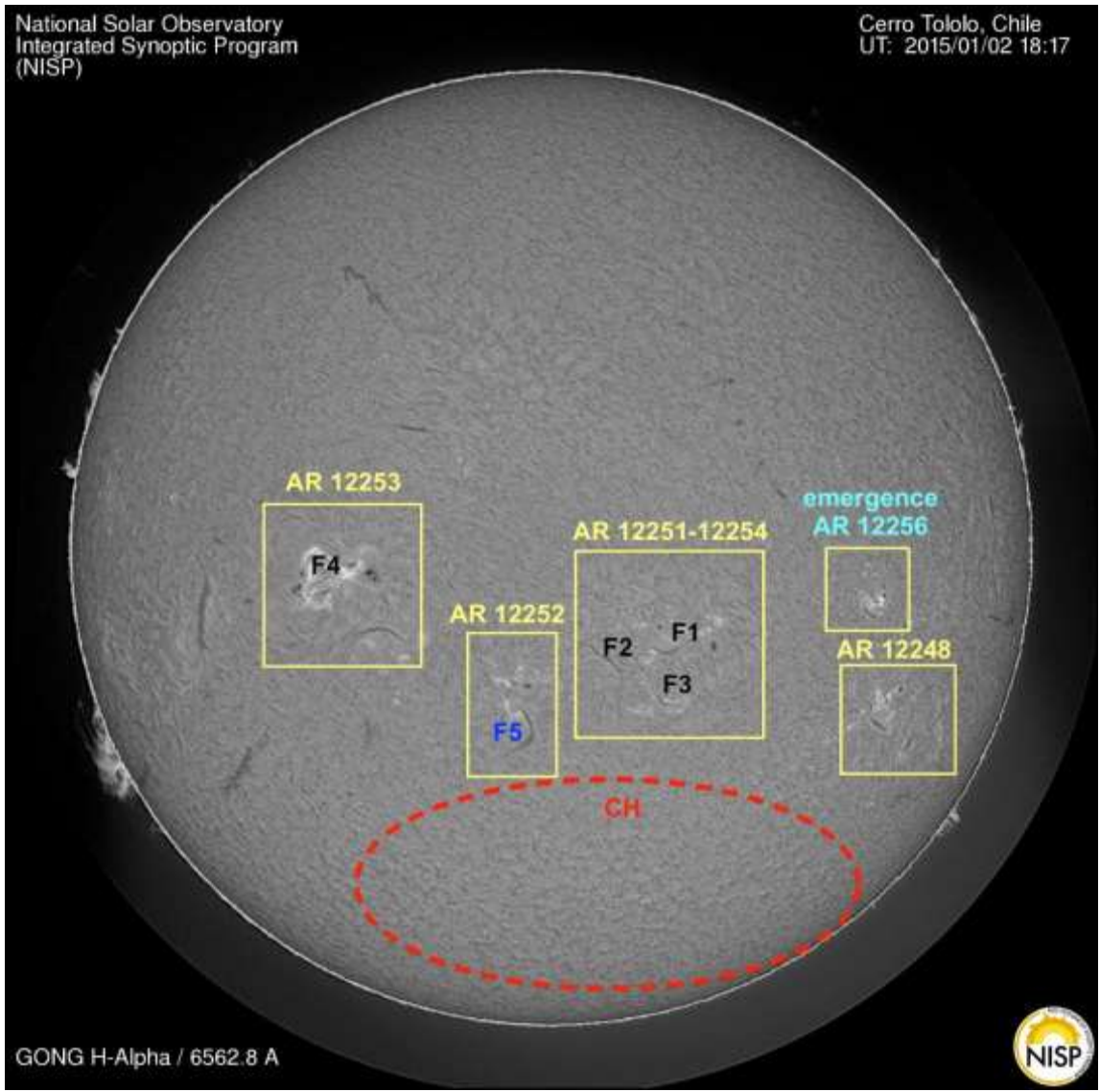

FIGURE 4

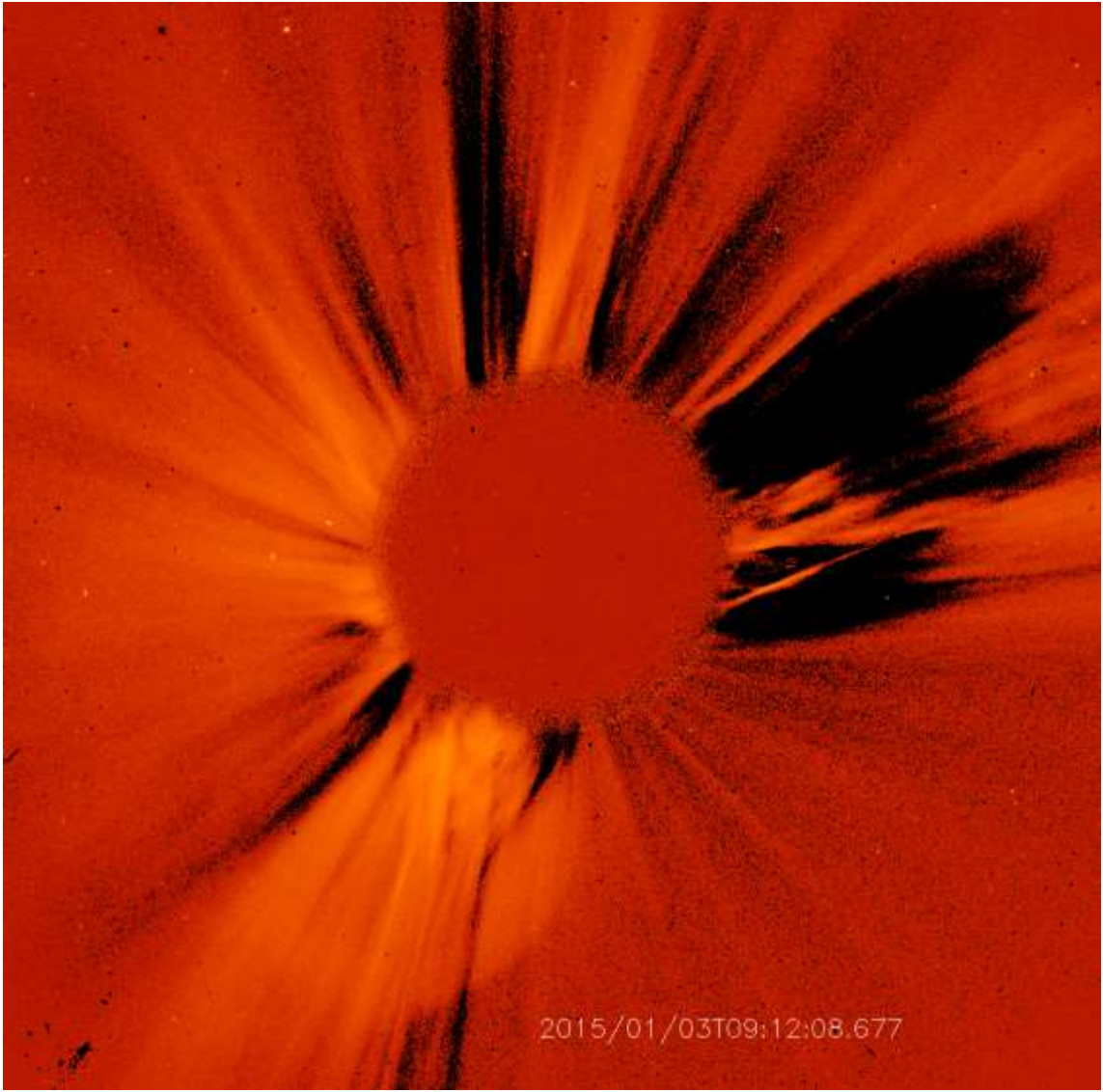

# FIGURE 5

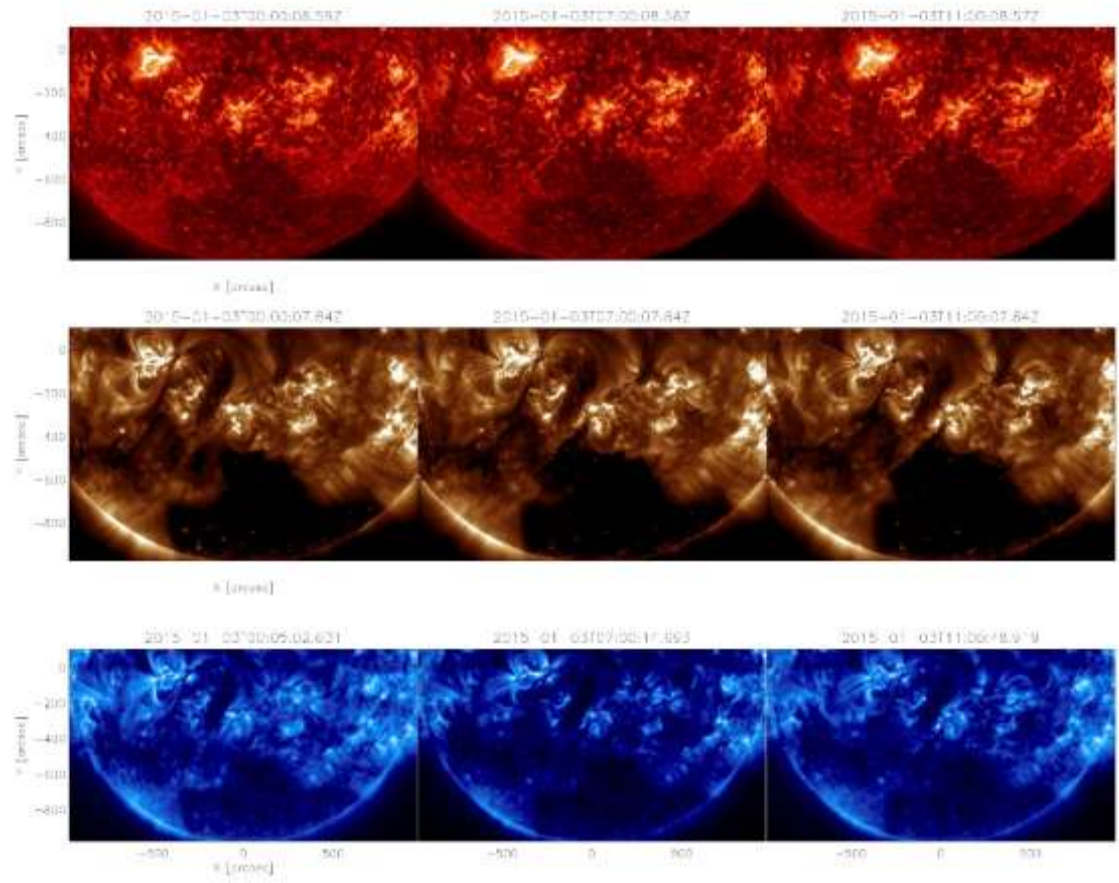

FIGURE 6

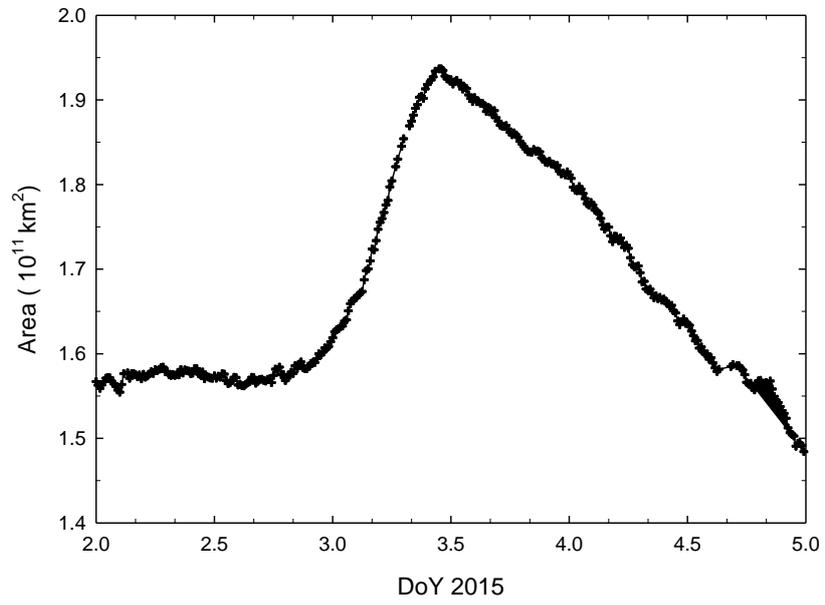

FIGURE 7

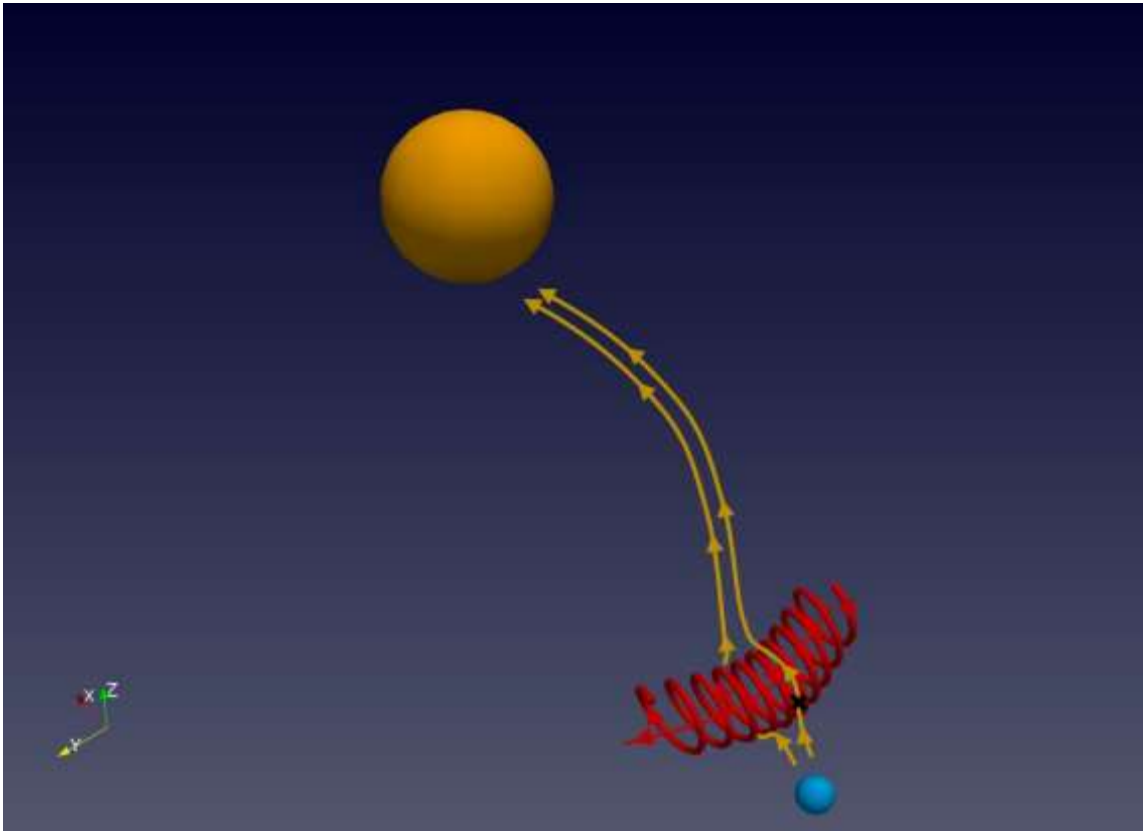